\documentclass[12pt]{article}

\newcommand{\wdot}{\dot{\omega}}
\newcommand{\phidot}{\dot{\phi}}
\newcommand{\sigdot}{\dot{\sigma}}
\newcommand{\wddot}{\ddot{\omega}}

\newcommand{\sgn}{\mbox{Sign}}

\newcommand{\eq}[1]{equation~(\ref{#1})}
\newcommand{\eqs}[2]{equations~(\ref{#1}) and~(\ref{#2})}
\newcommand{\eqto}[2]{equations~(\ref{#1}) to~(\ref{#2})}

\newcommand{\Rgb}{R^{2}_{\mbox{\tiny GB}}}

\begin{document}
\thispagestyle{empty}

\mbox{}

{\raggedleft \footnotesize
% hep-th/9605173 \\
% WU-AP/58/96\\
}

\vspace{3mm}

\begin{center} 	{\Large\bf\expandafter{
      One-Loop Superstring Cosmology \\
             and the \\
       Non-Singular Universe \\}}
\end{center}

\vspace{7mm}

\begin{center}
{\Large Richard Easther}
\footnote{ easther@cfi.waseda.ac.jp}
\medskip

		{\Large Kei-ichi Maeda}
\footnote{ maeda@cfi.waseda.ac.jp}

\bigskip
             Department of
Physics, \\ Waseda University,
 3-4-1  Okubo, Shinjuku-ku, \\ Tokyo, Japan.
\end{center}

\vspace{5mm}

\setcounter{footnote}{0}
\setcounter{page}{0}

\section*{Abstract}

We study the cosmological implications of the one-loop terms in the
string expansion.  In particular, we find non-singular solutions which
interpolate between a contracting universe and an expanding universe,
and show that these solutions provide a mechanism for removing the
initial conditions problem peculiar to spatially closed FRW
cosmologies.  In addition, we perform numerical calculations to show
that the non-singular cosmologies do not require a careful choice of
initial conditions, and estimate the likely magnitude of higher order
terms in the string expansion.

\newpage
\section{Introduction}

The study of cosmology leads inexorably to an epoch in which the
energy density of the universe approaches the Planck scale.
Superstring theory is the leading candidate for a description of
physics at the Planck scale, but has not developed to the point where
the full theory may be employed to construct a detailed cosmology.
Despite this, the generic properties of superstring models can be
extracted and their cosmological consequences investigated.  In doing
so, the cosmologist simultaneously hopes to find mechanisms for
resolving the various problems that remain after the introduction of
the inflationary scenario, and to test string theory in the high
energy laboratory provided by the early universe.

Two distinct approaches have been taken to the study of string
cosmology.  Firstly, many cosmologists approximate the full string
theory with the first terms of the perturbative string expansion
\cite{FradkinET1985a,CallanET1985a,Lovelace1986a}.
\nocite{BailinET1985b,Maeda1987a,AntoniadisET1988a,AntoniadisET1989a,%
LiddleET1989a,Mueller1990a,AntoniadisET1991b,Veneziano1991a,Tseytlin1991a,%
CasasET1991b,GarciaBellidoET1992b,Tseytlin1992b,TseytlinET1992a,%
BrusteinET1994b,BehrndtET1994a,GoldwirthET1993a,CopelandET1994b,EastherET1995a}
%
% -line above for the benefit of bibtex!
%
The non-perturbative features of string theory are implicitly
discarded, so this approach breaks down at the highest energies.
However, since Einstein gravity and conventional particle physics
agree well with experiment, a successful superstring theory must
reduce to these theories in the low energy limit.  Consequently, it is
reasonable to expect that at energy scales when stringy effects first
become readily apparent it will be possible to approximate the full
string theory by an effective theory comprising the usual
Einstein-Hilbert action together with higher order corrections.  In
particular, much of the effort to date has focussed upon the lowest
order, or tree-level action \cite{BailinET1985b}-\cite{EastherET1995a}.

Gasperini, Veneziano and others have adopted an alternative strategy,
which they have dubbed the ``pre-big-bang scenario''
\cite{GasperiniET1992b,GasperiniET1994a}.  This is a non-singular
cosmology, and it attempts to alleviate many of the problems
associated with the ``standard'' big bang by appealing to the
symmetries of the full superstring action, rather than a perturbative
limit.  The fundamental requirement for such a model is the existence
of non-singular or ``branch-changing'' \cite{Veneziano1991a,BrusteinET1994b}
solutions which smoothly
interpolate between a contracting universe and an expanding one
without passing through a singularity.

The term branch-change originally referred to a non-singular
transition between two solutions of a simple tree-level string
cosmology.  The solutions are distinguished by a sign choice and are
related to one another by a duality transformation.  Moreover, this
discussion was largely based in the (tree-level) string frame, which
is related to the usual Einstein frame of general relativity by a
conformal transformation.  A universe which is expanding in the string
frame may be contracting in the Einstein frame, and it is therefore
dangerous to simply identify the two branches with ``expansion'' and
``contraction''.  However, in general the $(-)$ branch evolves away
from a past singularity, while the $(+)$ branch represents evolution
towards a future singularity \cite{KaloperET1995a}.  The type of
branch-change required by the pre-big-bang scenario is a non-singular
transition from the $(+)$ to the $(-)$ branch.  The converse, a
solution which moves from $(-)$ to $(+)$ is entirely unremarkable, as
it corresponds to the generic evolution of a spatially closed FRW
universe.

Since string theory is expected to remove, or at least tame, the
infinities inherent in other models of fundamental physics, it is
reasonable to expect that it will remove the singularity represented
by the big bang in conventional cosmology.  At tree-level in the
superstring action, though, the possibility of classical
branch-changing solutions to the classical equations of motion has
been virtually excluded
\cite{EastherET1995a,KaloperET1995a,KaloperET1995b}, although quantum
cosmology may provide a mechanism for allowing a transition between
the two tree-level branches \cite{GasperiniET1996a,Lidsey1996a}.
However, the tree-level action is only the lowest order term in the
full string loop expansion, so consequently in this paper we explore
the cosmological solutions that arise when one loop contributions from
the dilaton and modulus terms are included.  We show that in this
case, the low-energy limit of string theory naturally leads to a
``bouncing universe'', in contrast to the tree-level limit.

Previously, Antoniadis, Rizos and Tamvakis
\cite{AntoniadisET1993a} have studied the equations of motion
derived from the same action with a spatially flat background and
found some non-singular solutions.  However, in this case the scale
factor increases monotonically from a (non-zero) constant value, which
differs from the branch-changing solutions envisaged by the
pre-big-bang scenario, as there is no transition from contraction to
expansion.  We extend this system to the case where the spatial
hypersurfaces are allowed to have non-zero curvature.  In doing so we
find ``bouncing'' solutions and investigate the range of initial
conditions over which they occur.  Since we do not directly employ the
duality symmetries of the superstring action in our analysis, we do
not explicitly identify these bouncing solutions with the
branch-changing solutions of the pre-big-bang scenario.  However, from
a phenomenological perspective, a bouncing solution resembles a
successful branch-change, as in both cases the universe is apparently
evolving towards a singularity in the distant past and away from it in
the distant future.

We also consider the particular initial conditions problem faced by a
spatially closed FRW universe, which in the absence of fine-tuning
will typically have a lifetime on the order of the Planck scale
\cite{KolbBK1}.  This problem persists in the presence of inflation:
although inflation can begin at the Planck scale there is no guarantee
that it will do so.  Here we show that the contribution from the
one-loop terms can allow a closed universe to grow arbitrarily large,
ensuring that the universe will survive long enough for inflation to
begin.

Finally, we numerically integrate the equations of motion for a wide
range of initial conditions, to demonstrate that the bouncing
solutions do not require a highly restrictive choice of parameters.
We also argue on the basis of dimensional analysis that many of the
bouncing solutions do not evolve into a region where the
perturbative expansion is likely to break down, which makes it plausible
that these solutions will persist if higher loop terms are added to
the action.

\section{Action and Equations of Motion}

We take as our starting point the one-loop superstring action
\cite{AntoniadisET1993a,AntoniadisET1992a,AntoniadisET1992b},
\begin{equation}
S = \int{d^4 x \sqrt{-g}\left\{
   \frac{1}{2} R - \frac{1}{4}(D \phi)^2 - \frac{3}{4}(D \sigma)^2
   + \frac{1}{16}\left[\lambda e^\phi - \delta\xi(\sigma)\right]
  \Rgb \right\}}, \label{action}
\end{equation}
which contains contributions from the Ricci scalar $R$, the dilaton
$\phi$ and a modulus field $\sigma$.  The Gauss-Bonnet combination,
$\Rgb$, is
\begin{equation}
\Rgb = R_{\mu\nu\kappa\lambda}R^{\mu\nu\kappa\lambda} -
      4R_{\mu\nu}R^{\mu\nu} + R^2.
\end{equation}
Terms which appear at one loop in the string expansion but vanish when
the metric has the Robertson-Walker form have been dropped from the
above action.  We have adopted the same conventions and notation as
Antoniadis, Rizos and Tamvakis with the exception of the metric
signature, which we have set to $(-,+,+,+)$, and our units correspond
to the gravitational constant, $G =1/8\pi$.  The coefficient $\lambda$
is positive and determined by the four dimensional string coupling.
The sign of $\delta$ is determined by the relative numbers of chiral,
vector and spin-$3/2$ massless supermultiplets, and is proportional to
the four dimensional trace anomaly of the $N=2$ sector of the theory.
It will be important that $\delta$ can take both positive and negative
values.  The potential, $\xi(\sigma)$ is defined in terms of the
Dedekind $\eta$ function,
\begin{equation}
 \xi(\sigma) = \ln{\left[ 2 e^{\sigma} \eta^{4}(ie^{\sigma})\right]}
\end{equation}
where $\eta$ is \cite{ErdelyiBK3}
\begin{equation}
\eta(\tau) = q^{1/12}\prod_{n=1}^{\infty}(1-q^{2n}),
\qquad q = e^{{i\pi\tau}}.
\end{equation}
Anticipating that first and second derivatives of $\xi$ will appear in
the equations of motion, we note that
\begin{equation}
\xi_{\sigma}(\sigma) = 1 - \frac{\pi e^{\sigma}}{3} +
  8\pi e^{\sigma} \sum_{n=1}^{\infty}
  \frac{n e^{-2n\pi e^{\sigma}}}{1 - e^{-2n\pi e^{\sigma}}}
  \label{xisigma1}
\end{equation}
where the subscript denotes differentiation with respect to $\sigma$.
Despite its appearance, this is an odd function of $\sigma$.
Furthermore, for large $|\sigma|$
\begin{equation}
\xi_{\sigma} \approx -\frac{2\pi}{3} \sinh{(\sigma)}  \label{xisigma2}
\end{equation}
closely approximates the exact expression, \eq{xisigma1}, and is also
antisymmetric under $\sigma \rightarrow -\sigma$ .  The accuracy of
this approximation for $\sigma \approx 0$ could be improved by adding
terms of the form $c_{n}\sigma^{n}$ for $n = 1,3,5\ldots$ (since
$\xi_{\sigma}$ is odd, even powers of $\sigma$ will not appear) with
the $c_{n}$ chosen so that the approximation reproduces the first few
terms in the Taylor expansion of $\xi_{\sigma}$ about $\sigma = 0$.
However, in practice adding these corrections does not alter the
qualitative properties of the solutions obtained numerically, so for
simplicity we have worked exclusively with \eq{xisigma2}.

Previously Antoniadis, Rizos and Tamvakis  examined the cosmological
solutions for this system that have a spatially flat Robertson
Walker metric.  We extend their work to include the possibility
that the spatial hypersurfaces have non-zero curvature, and so the
appropriate ansatz for the line element is
\begin{equation}
ds^{2} = -dt^{2} + e^{2\omega(t)}
  \left[ \frac{1}{1-kr^{2}} + r^{2} \left( d\theta^{2} +
  \sin^{2}\theta d\phi^{2}\right) \right].
\end{equation}
We therefore derive the equations of motion
\begin{eqnarray}
&3\left(\dot{\omega}^2 +  k e^{-2\omega}  \right)
    \left( 1 + 8 \dot{f}\dot{\omega} \right)  - \frac{1}{4} \dot{\phi}^2
  - \frac{3}{4} \dot{\sigma}^2 = 0& \label{const1} \\
& 2\left( \ddot{\omega} + \dot{\omega}^2 \right)
    \left(1 + 8 \dot{f} \dot{\omega}\right)
 + \left( \dot{\omega}^{2} +  k e^{-2\omega} \right)
       \left(1 + 8 \dot{f} \ddot{\omega}\right)   +
\frac{1}{4} \dot{\phi}^2  + \frac{3}{4} \dot{\sigma}^2    =0, &
    \label{wdd1} \\
& \ddot{\phi} + 3\dot{\omega}\dot{\phi} - 2\frac{d f}{d\phi} \Rgb =0,&
  \label{pdd1} \\
& \ddot{\sigma} + 3\dot{\omega}\dot{\sigma} -
    \frac{2}{3}\frac{d f}{d\sigma} \Rgb =0, &  \label{sdd1}
\end{eqnarray}
where a dot denotes differentiation with respect to $t$.  The
Gauss-Bonnet term is
\begin{equation}
\Rgb = 24\left(\ddot{\omega} + \dot{\omega}^2\right)
   \left(\dot{\omega}^2 + \frac{k}{e^{2\omega}}\right)
\end{equation}
and $f$ is defined to be
\begin{equation}
f = \frac{1}{16} \left[\lambda e^\phi  - \delta \xi(\sigma)\right].
\end{equation}
If $f$ vanishes, the system reduces to two free, minimally coupled,
scalar fields in a Robertson-Walker universe.

The equations above consist of three second order equations and a
constraint, giving a total of five degrees of freedom.  In order to
facilitate the analysis we isolate each of the second derivative terms
(remembering that $\ddot{f}$ implicitly contains $\ddot{\phi}$ and
$\ddot{\sigma}$), on the left hand side, giving the following system
\begin{eqnarray}
\ddot{\omega} &=&  -\dot{\omega}^2
- \left( \dot{\omega}^2 + \frac{k}{e^{2\omega}}\right) \chi,
 	  \label{wdd2} \\
\ddot{\phi} &=& -3\dot{\omega}\dot{\phi}  -3\lambda e^\phi
  \left( \dot{\omega}^2 + \frac{k}{e^{2\omega}}\right)^2 \chi,
  \label{pdd2} \\
 \ddot{\sigma} &=& -3\dot{\omega}\dot{\sigma} +\delta \xi_{\sigma}
 \left( \dot{\omega}^2 + \frac{k}{e^{2\omega}} \right)^2 \chi,
       \label{sdd2}
\end{eqnarray}
where
\begin{equation}
\chi = \frac{8 + \lambda\dot{\phi}^2 e^\phi - \delta \dot{\sigma}^2
   \xi_{\sigma\sigma} }{
   4 + 2(\lambda\dot{\phi}e^\phi - \delta \dot{\sigma}\xi_{\sigma})\dot{\omega}
  + (\dot{\omega}^2 + k e^{-2\omega})^2 (3\lambda^2 e^{2\phi} +
  \delta^2\xi_{\sigma}^2)}. \label{chi}
\end{equation}
If we wish, we can eliminate any one of the variables by inserting
the constraint. However, it will be more convenient to work with the
equations as they are given above, both in the following section when
we consider the asymptotic form of the solutions and in Section~4,
where the constraint will allow us to the check the accuracy of our
numerical solutions.

\section{Asymptotic Solutions}

In general this system of equations must be solved numerically.
However, considerable insight into the cosmological properties of this
model can be gained from analytic considerations alone.  In
particular, the existence of a bouncing solution requires that the
scale factor to pass through a (non-zero) minimum value and be
singularity free.  Hence we begin our investigation by considering the
possible extrema of $\omega$, $\phi$ and $\sigma$.

When the scale factor, $a = e^{\omega}$, passes through a local
minimum, its second derivative, $\ddot{a}$, must be positive.  When
$k=0$ the constraint, \eq{const1}, with $\wdot = 0$ requires $\sigdot
= \phidot = 0$ as well.  This is an exact static solution to the
equations of motion where the values of $\omega$, $\sigma$ and $\phi$
are all arbitrary constants.  The non-singular solutions of
Antoniadis, Rizos and Tamvakis can be regarded as the consequence of
making a small deviation from this solution in the distant past.  If
the spatial hypersurfaces have negative curvature, $k=-1$ and the
constraint cannot be satisfied when $\dot{\omega} = 0$.  Consequently,
the scale factor is again monotonic, but the constant solution no
longer exists.

When the spatial hypersurfaces have positive curvature, so that $k=1$,
the constraint can be satisfied with $\dot{\omega} = 0$ and
$\dot{\phi}$, $\dot{\sigma} \neq 0$.  Thus, in this case, the scale
factor may pass through an extremal value.  Since the first two terms in
the denominator of $\chi$ are equal to $4(1 + 8\dot{f}\wdot)$ it
follows from the constraint that if $k=1$ the denominator of $\chi$ is
positive, so
\begin{equation}
\sgn[\left. \wddot \right|_{\wdot=0}] =
   \sgn[-(8 + \lambda\dot{\phi}^2 e^\phi - \delta
   \dot{\sigma}^2\xi_{\sigma\sigma})]
\end{equation}
While $\lambda$ is physically restricted to positive values, $\delta$
can take any real value.  Since $\xi_{\sigma\sigma}(\sigma) < 0$ for
all $\sigma$, if $\delta \geq 0$, $\wddot$ will be negative at an
extremum of $\omega$, but can take either sign if $\delta < 0$.  This
immediately proves the existence of solutions for which $k=1$ and
$\delta < 0$ where the scale factor possesses a local minimum, which
is a prerequisite for a successful bounce.  However, this does not
establish the existence of globally non-singular solutions, since it
does not guarantee that such a solution will always be non-singular.
For instance, when the tree level action contains contributions from
the spatial curvature and an axion the Einstein frame scale factor
may pass through several local minima, but all solutions contain at
least one singularity \cite{EastherET1995a}.

Now consider the asymptotic form of solutions to the equations of
motion.  We focus on $k=1$, which is the only choice that can lead to
a bouncing solution and investigate the properties of a universe which
is expanding at large positive times
\begin{eqnarray}
\mbox{Type I} & \ddot{a}(t) > \epsilon,&   t > T, \nonumber \\
\mbox{Type II} & \ddot{a}(t) < -\epsilon, &  t > T,  \\
\mbox{Type III} & \ddot{a}(t) \rightarrow 0, &  t \rightarrow
\infty,   \nonumber
\end{eqnarray}
where $\epsilon$ and $T$ are both positive numbers. A solution which
is contracting at large, negative times corresponds to the
time-reverse of this case and does not need to be examined separately.

Since the defining condition for inflation is that $\ddot{a} > 0$
\cite{AbbottET1984a}, a solution of Type~I inflates forever and so
the curvature terms will be negligible at late times.  However, this
would imply the existence of solutions for $k=0$ that are inflationary
at late times, in contradiction with the results of Antoniadis, Rizos
and Tamvakis, so it follows that there are no expanding, asymptotic
solutions with the form of Type~I.  Conversely, in the case of
Type~II, the curvature term will dominate the $\wdot^{2}$ term where
they appear together in the equations of motion, and the scale factor
will eventually pass through a maximum.  Thus there are no solutions
which expand indefinitely where the scale factor has the generic form
of Type~II.  Note that if the scale factor does evolve through a local
maximum, it may either make a non-singular transition to a subsequent
stage of expansion or collapse to a singularity.

Finally, consider possible asymptotic solutions of Type~III.  Since
the Gauss-Bonnet term vanishes if $\ddot{a}$ is identically zero,
choosing $a(t) = a_{0}t$ is not a viable candidate for a late time
solution that involves a non-trivial contribution from the one-loop
terms.  Consequently, we need to consider solutions with an asymptotic
form like $a= e^{\omega} \rightarrow a_{1}t + a_{2}t (\ln{t})^{m}$
which are of Type~III but allow for the possibility of a non-zero
Gauss-Bonnet combination.

Dropping the contributions from the dilaton terms, make the following
substitution, where $\tau = \ln{t}$
\begin{eqnarray}
  A(t) &=& \frac{a(t)}{t} = a_{1} + a_{2} \tau^{m} \label{atry} \\
  s(t) &=& \frac{e^{\sigma}}{t^{2}} = s_{1} + s_{2} \tau^{n}
  \label{stry}
\end{eqnarray}
which allows us to write \eqs{const1}{sdd1} as
\begin{equation}
 \left[ \left(1 +  {{A'}\over{A}} \right)^{2} + \frac{1}{A^{2}}\right]
   \left[4 + 2 \Delta (2s + s') \left(1 + \frac{A'}{A}\right)\right]
   - \left(2 + \frac{s'}{s}\right) = 0
\end{equation}
\begin{equation}
  {{s''} \over {s}}   =   -\left(2 + 3\frac{A'}{A}\right)
  \left(2 + \frac{s'}{s}\right) +
  \Delta s \left( \frac{A''}{A} + \frac{A'}{A} \right)
  \left[ \left(1 + \frac{A'}{A}\right)^{2} + \frac{1}{A^{2}}\right]
\end{equation}
where $\Delta = \pi \delta/3$, a dash denotes differentiation with
respect to $\tau$ and we have assumed that $\sigma \gg 1$
so that $\xi_{\sigma} \propto e^{\sigma}$. This involves no loss of
generality, since the full equations of motion are invariant under the
transformation $\sigma \rightarrow -\sigma$.

With the ansatz, \eqs{atry}{stry}, we can systematically expand in
powers of $\tau$.  There are a number of subcases, corresponding to
$a_{1,2}$, $s_{1,2}$, $m$ and $n$ being either positive or negative,
subject to the overall requirement that $A(t)$ and $S(t)$ are
positive.  For $k=1$ and $\delta < 0$ (restricting attention to the
possible bouncing solutions) we can show that the only asymptotic
solution of this type is:
\begin{eqnarray}
A &=& \frac{1}{\sqrt{8 \tau}}, \label{atry2} \\
s &=& \frac{-1}{\Delta} \left(1 + \frac{5}{8 \tau} \right) \label{stry2} .
\end{eqnarray}
If $\phi \ll |\sigma|$ and the scale factor is given approximately by
\eq{atry2} then the one-loop terms have a negligible effect in
\eq{pdd1} and it follows that in the asymptotic regime the dilaton
$\phi$ is effectively constant.  As written above, this asymptotic
form describes an expanding universe as $t \rightarrow \infty$.  A
universe which is contracting as $t \rightarrow -\infty$ corresponds
to the time reversal of this solution.

We now have the ingredients we need to construct a universe that can
be arbitrarily large in the distant past, contract to a non-zero
minimum size and then make a smooth transition to expansion, after
which it may grow indefinitely.  A specific numerical solution of the
equations of motion which exhibits these properties is shown in
Figure~1.  The key ingredient in these solutions is the presence of
the modulus field $\sigma$ and its associated one-loop terms.  If
either $\delta \geq 0$, or the one-loop terms are simply absent from
the action, then the only type of extremum that the scale factor,
$a(t)$, can possess is a local maximum, and a bounce does not
occur.

An analogous result can be found for $\delta > 0$, but in this case no
bounce is possible.  Furthermore, any asymptotic solution to the scale
factor-modulus system for positive $\delta$ will be equivalent to a
similar solution to the scale factor-dilaton system in which $\phi
\rightarrow \infty$, although this implies that the system evolves
into the strong-coupling region and such a solution is therefore
unphysical.

Solutions of this type extend the previous work of Antoniadis, Rizos
and Tamvakis, further demonstrating that the singularity problem is
less acute in cosmological models based on the one-loop string action
than at tree-level or in classical, Einstein gravity.  This provides
empirical support for the hope that a fully non-perturbative string
cosmology will resolve the singularity problem entirely.

Physically, the key ingredient in the non-singular solutions is the
one-loop term that couples the Gauss-Bonnet combination and the
modulus; without this there are no non-singular solutions.  However,
the sign of this coupling is determined by $\delta$ and it is only for
$\delta < 0$ that the one-loop modulus terms can cause a bounce.
Thus, while have established that adding one-loop terms to the low
energy string action introduces non-trivial non-singular solutions and
significantly softens the conclusions reached at tree-level, these
corrections do not imply a complete absence of singularities.

The new class of non-singular cosmologies found here shows that string
theory has the ability to resolve the curvature problem typically
associated with a closed, FRW universe, which has a typical lifetime
(and maximum size) on the order of the Planck scale.  Inflation can
easily solve the analogous problem in an open universe, which will
expand indefinitely, whereas in a spatially closed universe inflation
is typically only successful if it begins at the Planck scale.
However, we have shown that the presence of the one-loop modulus terms
permit a closed FRW universe to expand indefinitely.  This is not
sufficient to solve the flatness problem, but these solutions provide
a mechanism for ensuring that the universe lasts long enough and the
energy density becomes low enough for inflation to begin without the
need for any additional constraints in a $k=1$ FRW universe.  This is
a property of the asymptotic form of the solutions, and solutions
which contain an initial singularity may still expand indefinitely.
Furthermore, the analogous asymptotic form for $\delta > 0$ also
allows indefinite expansion, even though a non-singular, bouncing
solution is impossible in this case.

Solutions of the type displayed in Figure~1 strongly resemble the
type of branch-change envisaged by the pre-big-bang scenario.  Unlike the
non-singular solutions that exist when $k=0$, the scale factor is not
monotonic.  As $t \rightarrow \pm\infty$ the dilaton is fixed, and the
Einstein and string frame scale factors differ only by a
multiplicative constant.  Hence there is no ambiguity in associating a
solution which is contracting in the string frame at early times to
one that is contracting in the Einstein frame, as there might be if
the dilaton were evolving rapidly.

We can also gain some insight into the behavior of the dilaton and
modulus fields.  Consider \eqs{pdd2}{sdd2}, and recall that the
denominator of $\chi$ is positive for $k=1$, so
\begin{eqnarray}
\sgn[\left. \ddot{\phi} \right|_{\phidot=0}] &=&
    \sgn\left[ -(8   - \delta \dot{\sigma}^2\xi_{\sigma\sigma})
    \right], \\
\sgn[\left. \ddot{\sigma} \right|_{\sigdot=0}] &=&
    \sgn\left[ \delta \xi_{\sigma}(\sigma) \right],
\end{eqnarray}
since $\lambda >0$ on physical grounds.  Thus when $\delta < 0$,
which is the case we are most interested in, the dilaton can possess
both maxima and minima and, at least in principle, may undergo several
oscillations before the solution becomes established in one of the
asymptotic regimes.  However, any extremal value of $\sigma$ when
$\sigma < 0$ is a local maximum, and any extremal value when $\sigma >
0$ is a local minimum.  Thus when $\delta < 0$, $\sigma$ can have at
most one extremum as it evolves from one asymptotic regime to the
other.

When the scale factor $a$ passes through its minimum value, $\ddot{a}
> 0$.  As this is the minimal requirement for the existence of
inflation, it follows that in a pedantic sense a bouncing solution is
also inflationary.  Obviously, the astrophysical constraints that a
successful inflationary model must satisfy are much more demanding
than simply requiring that the scale factor undergo positive
acceleration.  At late times the asymptotic form of the solution
approaches the borderline condition between inflationary and regular
growth, as $\ddot{a} \rightarrow 0$, which resembles the coasting
solutions that have been discussed in the context of standard
inflationary cosmology \cite{EllisET1991c}.

In the next section we use numerical techniques to investigate the
generality of these non-singular solutions.  However, note that while
the scale factor is monotonic if $k=-1$, there is a simple exact
solution
\begin{eqnarray}
\omega &=& \omega_{0} + \ln{t}, \\
\phi &=& \phi_{0}, \\
\sigma &=& \sigma_{0},
\end{eqnarray}
which describes an empty, curvature dominated universe, and $\ddot{a}
\equiv 0$.  However, unlike the linearly expanding solution found
for $k=1$, this solution holds when $\lambda$ and $\delta$ are zero
and the Gauss-Bonnet combination will vanish exactly, so it is not
related to the one-loop terms in the string action.

\section{Numerical Results}

In the previous section we displayed a particular example of a
non-singular universe, associated with the one-loop terms in the
perturbative expansion of the superstring action.  In this section we
estimate of the likely impact of higher loop terms and investigate the
range of initial conditions that give rise to a non-singular universe.

While it is simple to determine which solutions are singular and which
are non-singular, the more qualitative distinction between a solution
which comes ``close'' to the Planck scale, and one where the bounce
occurs at a considerably lower energy scale is obviously important.
One way to quantify this is through the maximum values of the scalar
quantities that can be constructed from the metric curvature, such as
$R$, $R_{\mu\nu}R^{\mu\nu}$,
$R_{\mu\nu\kappa\lambda}R^{\mu\nu\kappa\lambda}$ or various higher
order combinations. In particular,  Brandenberger, Mukhanov and others
\cite{MukhanovET1992a,BrandenbergerET1993a} have constructed
models that ensure these quantities remain sub-Planckian for all
homogenous and isotropic cosmological solutions.  The action
considered by us does not have this property since it does admit some
singular solutions, for which the curvature invariants will exceed any
given finite value.  However, the curvature invariants derived from
the non-singular solutions are typically sub-Planckian.  As a specific
example, the value of $R_{\mu\nu\kappa\lambda}R^{\mu\nu\kappa\lambda}$
for the solution of Figure~1 is plotted in Figure~2.

While it is reassuring that the curvature terms do not exceed the
Planck scale, the dilaton and modulus fields couple directly to
higher-order curvature terms, and if the impact of higher terms in the
perturbative expansion is to be small then these combinations must be
sub-Planckian as well.  Without a detailed calculation at two-loop
order and beyond, it is impossible to state definitively whether the
non-singular solutions found at one-loop will persist if higher order
terms are added to the action.  However, on dimensional grounds the
two-loop terms can be expected to be smaller than the one-loop
Gauss-Bonnet term provided that $\wdot$, $\phidot$ and $\sigdot$ do
not become significantly greater than unity (in Planckian units) while
$a = e^{\omega}$ remains less than unity.  Also, if $e^{\phi}$ (which
is effectively the loop expansion parameter) becomes large, we can
expect a significant contribution from higher-order terms.  Hence on
the basis of dimensional analysis alone we expect that the two-loop
terms will be at worst roughly equal to the one-loop terms if
\begin{equation}
\wdot, \quad \phidot, \quad \sigdot, \quad e^{\phi}, \quad 1/a < 1
\label{bound}
\end{equation}
at all times.  Note that this is typically a harsher constraint than
requiring that the curvature invariants do not exceed the Planck
scale.

Having established a heuristic criterion for the importance of higher
order terms, we also wish to establish whether bouncing solutions
require a special choice of initial conditions.  Given $\lambda$ and
$\delta$, a solution to the equations of motion is fully specified by
five initial conditions, the sixth being fixed by the constraint.  The
set of points in ``initial conditions space'' for which the system
evolves towards a particular attractor (that is, a particular
asymptotic solution) is known as the ``basin of attraction'' of the
corresponding attractor.  The full system of equations is five
dimensional, and therefore cannot be easily visualized using a phase
space approach.  Since the non-singular dynamics depends on the
behavior of the modulus field we could set $\lambda$ to zero (or drop
the dilaton entirely) and reduce the system to three dimensions.
Previously, however, Cornish and Levin \cite{CornishET1995a} have
examined the basins of attraction for two field inflationary models,
whose equations of motion are similar to those considered here, by
numerically integrating to find the asymptotic behavior for many
choices of initial conditions.  Using this technique we do not have to
restrict ourselves to a reduced system in which the dilaton is
trivial, and we can investigate the basin of attraction for solutions
where $\omega$ and $\sigma$ have the asymptotic form of
\eqs{atry}{stry} and the dilaton tends towards a constant.

In Figure~3 we display the asymptotic form of the solutions derived
from a sequence of two dimensional slices through the initial
conditions space, where each slice represents a $400 \times 400$ grid
of $\wdot_{0}$ and $\sigdot_{0}$ values, for five choices of
$\sigma_{0}$ between 0 and 2.  Since the equations of motion are
unchanged by the transformation $\sigma \rightarrow -\sigma$ there is
no need to carry out separate integrations for $\sigma_{0} < 0$.  The
integrations were carried out using the Bulirsch-Stoer method
\cite{PressBK1}, with the form of the equations of motion given by
\eqto{wdd2}{sdd2}, while the constraint was used to check the accuracy
of the numerical routines.

The intersection of the basins of attraction for a universe that is
non-singular as $t \rightarrow \pm \infty$ is shown in Figure~3.  The
intersection of these two attractors defines the region of initial
conditions space for which leads to a non-singular, bouncing
cosmology.  Since this is a substantial volume of the total range of
initial conditions, it follows that the universe does not have to be
``fine tuned'' in order to ensure that it is non-singular.
Furthermore, the results obtained here do not depend strongly on
$\phi_{0}$ or $\phidot_{0}$, or the magnitudes $\lambda$ and $\delta$.

As well as displaying the choices of initial data corresponding to a
bouncing solution, Figure~3 shows the subset of points for which the
higher-loop terms in the perturbative expansion are not expected to
make a substantial contribution.  We see that this in turn represents
a large subset of the total basin of attraction, suggesting that the
existence of non-singular solutions is not a quirk of the one-loop
action.  Thus it is clear that the non-singular solutions are in no
way special, and that a substantial subset of them are such that the
values of $a$, $\phi$, $\sigma$ and their corresponding velocities are
sufficiently small at all times to ensure that the action is unlikely
to be dominated by higher-loop corrections.

\section{Discussion}

In this paper we present explicit non-singular cosmological solutions
derived from the one-loop superstring action.  These solutions
represent a FRW spacetime where the scale factor makes a transition
from contraction to expansion, while remaining non-zero at all times.
As such, they resemble the type of branch-changing solution that is a
prerequisite for the successful implementation of the pre-big-bang
scenario, and demonstrate that although branch-changing does not occur
at tree level in the superstring action, non-singular ``bounce''
solutions do exist at one-loop.

While the non-singular solutions described here occur only when the
spatial hypersurfaces have positive curvature and the parameter
$\delta$ in the one-loop action is negative, our numerical
calculations establish that the non-singular solutions do not require
a careful choice of initial values for the scale factor, dilaton and
modulus fields.  Furthermore, while we have not made a detailed
analysis of the two-loop terms we expect that their contribution will
not dominate the one-loop terms, providing cause for cautious optimism
that these non-singular solutions will not vanish when higher loop
terms are incorporated into the action.

These solutions allow us to address the initial conditions problem
peculiar to the $k=1$ FRW universe, which typically has a lifetime on
the order of the Planck scale.  While GUT-scale inflation will solve
the other difficulties faced by the standard model of the big bang,
inflation can only prevent a closed FRW universe from recollapsing
if it commences at the Planck scale.  However, we have seen
here that for a wide range of initial conditions the equations of
motion derived from the one-loop string action permit a closed FRW
universe to expand indefinitely.  Consequently, this provides a string
motivated solution to the version of the curvature problem faced by a
$k=1$ universe without requiring that inflation commences at the
Planck scale.

The inflationary epoch cannot be described within the context of the
model discussed here, as we have not included matter fields in the
action that can drive inflation followed by a graceful exit to a
non-inflationary universe with the matter content we observe at low
energies.  Thus, while our solutions describe a universe which grows
arbitrarily large, we know that in practice the assumptions that
underpin the action we consider must break down at sufficiently
low energy scales.  Furthermore, the existence of solutions that
describe a closed FRW universe which grows arbitrarily large require
that the modulus field is also evolving continuously, which is made
possible by the fact the modulus field contains no explicit mass terms
at tree level.  In a realistic theory the moduli fields must become
massive after supersymmetry breaking, which puts a lower limit on the
validity of our action even in the absence of other matter fields and
non-perturbative effects.  Thus attempts to use the properties of
these solutions in a more realistic cosmological model must provide a
mechanism that will allow a ``graceful exit'' from the modulus
dominated expansion into an inflationary phase.  Furthermore, the late
time evolution of the dilaton must be small enough to ensure that the
constraints on the time variation of $G$, the Newtonian gravitational
constant, are obeyed.  We do not address this problem here, but note
that this difficulty afflicts almost all string inspired models, and
is not peculiar to the specific example considered here.

Physically, we have seen that the key difference between the one-loop
and tree-level actions that permits the existence of non-singular
solutions is the coupling between the modulus and the Gauss-Bonnet
combination.  However, the presence of this term alone cannot
guarantee a non-singular universe.  Firstly, while the parameter
$\delta$ can take on both positive and negative values, non-singular
solutions only arise when $\delta < 0$.  Secondly, even if $\delta$ is
negative, non-singular evolution is not guaranteed.  However, the
existence of non-singular solutions to the one-loop action represents
a significant improvement on the results found at tree-level and shows
that the perturbative limit of string theory can address the
singularity problem that is characteristic of the standard big-bang.

Generically, extensions to Einstein gravity consisting of combinations
of scalar fields and higher order curvature terms will not lead to a
singularity-free theory, but are more likely to make the existing
singularity problems worse.  Thus it is noteworthy that string theory
naturally leads to a low-energy action in which non-singular solutions
are possible.  More general theories have been considered by
Brandenberger and Mukhanov {\it et  al.}
\cite{MukhanovET1992a,BrandenbergerET1993a} who derive a higher order
gravitational action for which all homogeneous and isotropic solutions
are non-singular.  In doing so, they show that this requirement places
a strong constraint on the possible form of the action, in the absence
of fundamental scalar fields.  If fundamental scalar fields were
incorporated into their approach, it would be possible to investigate
whether models containing second order curvature invariants (such as
the Gauss-Bonnet combination) coupled to scalar fields generically
permit bouncing solutions and to what extent this is a special
property of the one-loop superstring action.  Conversely, Rizos and
Tamvakis \cite{RizosET1994a} show that with for a scalar field coupled
to the Gauss-Bonnet term with a spatially flat metric, non-singular
solutions are possible for a wider range of couplings than the
function $\xi(\sigma)$ that is derived from string theory, and it
would be of interest to extend their analysis to the $k=1$ case.

The fact that incorporating string loop contributions into the action
leads to a non-singular cosmology is reminiscent of previous attempts
to solve the singularity problem by including higher-order terms
derived from quantum corrections to the gravitational action
\cite{Nariai1971a,NariaiET1971a,Gurovich1977a}, which have long been
known to remove, or at least soften, the singularity associated with
the standard model of the big bang.

Superstring theory currently offers us a tantalizing glimpse of a
paradigm that will provide a unified description of fundamental
physics which modifies the predictions of general relativity in such a
way as to remove the initial singularity associated with the big bang.
At present it is impossible to know whether this promise will be
fulfilled, but there are strong indications that string theory
addresses most, if not all, of the problems of the conventional big
bang that can be traced back to the Planck scale.

\section*{Acknowledgments}

RE is supported by the JSPS, and a Grant-in-Aid for JSPS fellows
(0694194).  This work was partially supported by the Ministry of
Education, Science and Culture, Japan (06302021 and 06640412). We
wish to thank Neil Cornish for several useful conversations.

%
%\bibliography{massive,books}

\begin{thebibliography}{10}

\bibitem{FradkinET1985a}
E.~S. Fradkin and A.~A. Tseytlin, Phys. Lett. B {\bf 158},  316  (1985).

\bibitem{CallanET1985a}
C.~G. Callan, D. Friedan, E.~J. Martinec, and M.~J. Perry, Nucl. Phys. B {\bf
  262},  593  (1985).

\bibitem{Lovelace1986a}
C. Lovelace, Nucl. Phys. B {\bf 273},  413  (1986).

\bibitem{BailinET1985b}
D. Bailin and A. Love, Phys. Lett. B {\bf 163},  135  (1985).

\bibitem{Maeda1987a}
K. Maeda, Phys. Rev. D {\bf 35},  471  (1987).

\bibitem{AntoniadisET1988a}
I. Antoniadis, C. Bachas, J. Ellis, and D.~V. Nanopoulos, Phys. Lett. B {\bf
  211},  393  (1988).

\bibitem{AntoniadisET1989a}
I. Antoniadis, C. Bachas, J. Ellis, and D.~V. Nanopoulos, Nucl. Phys. B {\bf
  328},  117  (1989).

\bibitem{LiddleET1989a}
A.~R. Liddle, R.~G. Moorhouse, and A.~B. Henriques, Nucl. Phys. B {\bf 311},
  719  (1989).

\bibitem{Mueller1990a}
M. Mueller, Nucl. Phys. B {\bf 337},  37  (1990).

\bibitem{AntoniadisET1991b}
I. Antoniadis, C. Bachas, J. Ellis, and D.~V. Nanopoulos, Phys. Lett. B {\bf
  257},  278  (1991).

\bibitem{Veneziano1991a}
G. Veneziano, Phys. Lett. B {\bf 265},  287  (1991).

\bibitem{Tseytlin1991a}
A.~A. Tseytlin, Mod. Phys. Lett. A {\bf 6},  1721  (1991).

\bibitem{CasasET1991b}
J.~A. Casas, J. Garc\'{\i}a-Bellido, and M. Quir\'{o}s, Nucl. Phys. B {\bf
  361},  713  (1991).

\bibitem{GarciaBellidoET1992b}
J. Garc\'{\i}a-Bellido and M. Quir\'{o}s, Nucl. Phys. B {\bf 385},  558
  (1992).

\bibitem{Tseytlin1992b}
A.~A. Tseytlin, Class. Quantum Grav. {\bf 9},  979  (1992).

\bibitem{TseytlinET1992a}
A.~A. Tseytlin and C. Vafa, Nucl. Phys. B {\bf 372},  443  (1992).

\bibitem{BrusteinET1994b}
R. Brustein and G. Veneziano, Phys. Lett. B {\bf 329},  429  (1994).

\bibitem{BehrndtET1994a}
S. Behrndt and S. F\"{o}rste, Nucl. Phys. B {\bf 430},  441  (1994).

\bibitem{GoldwirthET1993a}
D.~S. Goldwirth and M.~J. Perry, Phys. Rev. D {\bf 49},  5019  (1994).

\bibitem{CopelandET1994b}
E.~J. Copeland, A. Lahiri, and D. Wands, Phys. Rev. D {\bf 50},  4868  (1994).

\bibitem{EastherET1995a}
R. Easther, K. Maeda, and D. Wands, Phys. Rev. D {\bf 53},  4247  (1996).

\bibitem{GasperiniET1992b}
M. Gasperini and G. Veneziano, Astropart. Phys. {\bf 1},  317  (1993).

\bibitem{GasperiniET1994a}
M. Gasperini and G. Veneziano, Phys. Rev. D {\bf 50},  2519  (1994).

\bibitem{KaloperET1995a}
N. Kaloper, R. Madden, and K.~A. Olive, Nucl. Phys. B {\bf 452},  677  (1995).

\bibitem{KaloperET1995b}
N. Kaloper, R. Madden, and K.~A. Olive, hep-th/9510117  (1995).

\bibitem{GasperiniET1996a}
M. Gasperini, J. Maharana, and G. Veneziano, hep-th/9602087  (1996).

\bibitem{Lidsey1996a}
J.~E. Lidsey, gr-qc/9605017  (1996).

\bibitem{AntoniadisET1993a}
I. Antoniadis, J. Rizos, and K. Tamvakis, Nucl. Phys. B {\bf 415},  497
  (1994).

\bibitem{KolbBK1}
E.~W. Kolb and M.~S. Turner, {\em The Early Universe}, Vol.~69 of {\em
  Frontiers in Physics} (Addison Wesley, Redwood City, California, 1990).

\bibitem{AntoniadisET1992a}
I. Antoniadis, E. Gava, and K.~S. Narain, Nucl. Phys. B {\bf 383},  93  (1992).

\bibitem{AntoniadisET1992b}
I. Antoniadis, E. Gava, and K.~S. Narain, Phys. Lett. B {\bf 283},  209
  (1992).

\bibitem{ErdelyiBK3}
A. Erd\'{e}lyi, W. Magnus, F. Oberhettinger, and F.~G. Tricomi, {\em Higher
  Transcendental Functions} (MacGraw-Hill, New York, 1955), Vol.~3.

\bibitem{AbbottET1984a}
L.~F. Abbott and M.~B. Wise, Nucl. Phys. B {\bf 244},  541  (1984).

\bibitem{EllisET1991c}
G.~F.~R. Ellis, D.~H. Lyth, and M. Miji\'{c}, Phys. Lett. B {\bf 271},  52
  (1991).

\bibitem{MukhanovET1992a}
V.~F. Mukhanov and R.~H. Brandenberger, Phys. Rev. Lett. {\bf 68},  1969
  (1992).

\bibitem{BrandenbergerET1993a}
R. Brandenberger, V. Mukhanov, and A. Sornborger, Phys. Rev. D {\bf 48},  1629
  (1993).

\bibitem{CornishET1995a}
N.~J. Cornish and J.~J. Levin, Phys. Rev. D {\bf 53},  3022  (1996).

\bibitem{PressBK1}
W.~H. Press, S.~A. Teukolsky, W.~T. Vettering, and B.~P. Flannery, {\em
  Numerical Recipes in FORTRAN}, 2 ed. (Cambridge University Press, Cambridge,
  1992).

\bibitem{RizosET1994a}
J. Rizos and K. Tamvakis, Phys. Lett. B {\bf 326},  57  (1994).

\bibitem{Nariai1971a}
H. Nariai, Prog. Theor. Phys. {\bf 46},  433  (1971).

\bibitem{NariaiET1971a}
H. Nariai and K. Tomita, Prog. Theor. Phys. {\bf 46},  776  (1971).

\bibitem{Gurovich1977a}
V.~T. Gurovich, Sov. Phys. JETP {\bf 46},  193  (1977).

\end{thebibliography}
%\bibliographystyle{prsty}

\newpage
\section*{Figure Captions}

\begin{figure}[htbp]
\caption[]{A particular non-singular solution is displayed, where
$\delta = -16\times 3/\pi$ and $\lambda = 1$.  The chosen initial
values are $\sigma_{0} = 0$, $\phi_{0} = -3$, $\dot{\phi}_{0} =
-5\times 10^{-4}$, $\dot{\sigma}_{0} = 0.08$ and $\wdot_{0} = 0.01$,
so that the constraint requires $\omega_{0} = 3.25114$, and all
quantities are expressed in Planckian units.  The scale factor passes
through a minimum value of approximately $10$, and the solution
approaches the asymptotic form of \eqs{atry2}{stry2} as $t \rightarrow
\pm\infty$.}
\end{figure}

\mbox{}

\begin{figure}[htbp]
\caption[]{The value of the scalar,
$R_{\mu\nu\kappa\lambda}R^{\mu\nu\kappa\lambda}$ is plotted for the
solution depicted in Fig.~1, demonstrating that it remains well below the
Planck scale at all times. Note the different scale on the time axis
in this plot.}
\end{figure}

\mbox{}

\begin{figure}[htbp]
\caption[]{Five different slices through the initial conditions space
are displayed, with $\phi_{0} = -3$, $\dot{\phi}_{0} = -0.005$,
$\lambda = 1$ and $\delta = -16 \times 3/ \pi$.  Note that not all
choices of initial conditions correspond to a universe with $k=1$.
The set of initial conditions leading to a non-singular universe
constitutes the basin of attraction for the bouncing solutions.  The
regions inside the solid lines correspond to solutions which do not
violate the approximate bounds on the validity of the one-loop
approximation,
\eq{bound}.}
\end{figure}

\end{document}